\documentstyle[prd,aps,epsfig]{revtex}

\def\Re{\mathop{\rm Re}}

\def\ubar{{\overline{u}}}
\def\dbar{{\overline{d}}}
\def\nubar{{\overline{\nu}}}
\def\ebar{{\overline{e}}}
\def\lsim{\mathrel{\rlap{\lower4pt\hbox{\hskip1pt$\sim$}}
    \raise1pt\hbox{$<$}}}         
\def\gsim{\mathrel{\rlap{\lower4pt\hbox{\hskip1pt$\sim$}}
    \raise1pt\hbox{$>$}}}         

\begin{document}
 
 

\title{Neutralino decay rates with explicit R-parity violation}
 
\author{Edward A. Baltz\footnote{E-mail address: eabaltz@wharton.berkeley.edu}}
\address{Department of Physics, University of California,\\
 601 Campbell Hall, Berkeley, CA 94720-3411}
\author{Paolo Gondolo\footnote{E-mail address:
    gondolo@mppmu.mpg.de}}
\address{Center for Particle Astrophysics, University of California,\\
301 Le Conte Hall, Berkeley, CA 94720-7304\\
{\rm and}\\
Max-Planck-Institut f\"{u}r Physik, Werner-Heisenberg-Institut,
\footnote{Present address.}\\
F\"ohringer Ring 6, 80805 M\"unchen, Germany.}

 
\maketitle

\vspace{.5in}
 
\begin{abstract}
  We compute the neutralino decay rate in the minimal supersymmetric standard
  model with the addition of explicit R-parity violation.  We include the
  complete squark and slepton mixing matrices, previously neglected, and we
  improve and correct published formulas.  These decays are relevant to
  accelerator and non-accelerator searches for R-parity violation, and are
  especially interesting in light of the reported high $Q^2$ anomaly at HERA.
\end{abstract}

\pacs{PACS numbers 12.60.-i, 12.60.Jv, 12.90.+b}

\twocolumn
\narrowtext
 
In the minimal supersymmetric standard model \cite{mssm}, a discrete symmetry
called R-parity is invoked to forbid gauge invariant lepton and baryon number
violating operators.  The R-parity of a particle is given by
$R_p=(-1)^{L+2S+3B}$, where $L$ and $B$ are the lepton and baryon numbers, and
$S$ is the spin.  Standard model fermions, Higgs bosons, and gauge bosons have 
$R_p=+1$, while their superpartners have $R_p=-1$.  This symmetry guarantees
the stability of the lightest superpartner (LSP).

There is no deep theoretical motivation for imposing R-parity, and it is an
interesting exercise to explore the phenomenology of R-parity violation
\cite{rp-viol}.  We introduce explicit R-parity violation by adding
\begin{equation}
W_{R\!\!\!/_p}=\lambda LLE^c+\lambda'LQD^c+\lambda''U^cD^cD^c
\end{equation}
to the superpotential.  The first two terms violate lepton number, and the
third violates baryon number.  The LSP can now decay into standard model
particles.

In models where supersymmetry is broken by supergravity, the LSP is usually the
lightest of the neutralinos, which are superpositions of the superpartners of
the neutral electroweak gauge bosons and the superpartners of the neutral Higgs
bosons.  R-parity violation allows the neutralino, which is a Majorana fermion,
to decay into three standard model fermions (see figure 1).  

Neutralino decays are relevant to accelerator searches for R-parity violation,
especially resonant squark production \cite{squark,dreiner}.  Astrophysical
neutralino decays are also of interest, and can put strong constraints on the
R-parity violating couplings \cite{us}.

It is interesting to note that neutralino decays may be relevant to the
reported high $Q^2$ anomaly at HERA \cite{hera}.  The anomaly is an excess of
events with a positron in the final state at high $Q^2$.  One interpretation
\cite{highq2} is resonant production of a squark $\tilde{u}$ in
$e^+d\rightarrow\tilde{u}$ due to an $LQD^c$ term in the superpotential.  The
squark may decay back to a positron via the R-parity violating operator or may
decay into $\chi^0 u$ and the neutralino $\chi^0$ then decay into a positron by
an R-parity violating interaction.  This scenario awaits confirmation, such as
from related charged current events \cite{cc}.

The calculations of neutralino decay rates into fermions are subtle because
they involve both Majorana fermions and fermion-number violating operators.  To
our knowledge, only one calculation is available in the literature
\cite{dreiner}, and it neglects sfermion mixing.  We improve this calculation
by including the complete sfermion mixings and we find some small but
significant differences.

The differential decay rate of the neutralino is given by a standard three-body
phase space factor multiplied by a squared amplitude which is averaged over the
initial neutralino spin and summed over the final fermion spins.  We denote
this spin sum by a primed summation symbol.
\begin{equation}
\frac{\partial^2\Gamma}{\partial x_1 \partial x_2} = { m_\chi \over 256 \pi^3 }
{\sum_{\rm spins}}' \left| {\cal M} \right|^2 \;.
\end{equation}
Here $x_{1,2}=2E_{1,2}/m_\chi$ and $E_{1,2}$ are two final state fermion
energies.

In the following we present the calculation of the spin averaged squared
amplitudes for neutralino decay.  We first consider the $U^c D^c D^c$ term in
the superpotential, followed by $LQD^c$ and finally $LLE^c$.

To calculate the $U^cD^cD^c$ decay, we first obtain the Feynman rule for the
R-parity violating vertex. Writing all indices explicitly the superpotential
reads
\begin{equation}
W_{UDD} = \epsilon^{\alpha\beta\gamma} \left[ \lambda''_{ijk} U^c_{i\alpha}
D^c_{j\beta} D^c_{k\gamma} \right],
\end{equation}
where $U^c_{i\alpha}$, $D^c_{j\beta}$ and $D^c_{k\gamma}$ are the superfields
of the right-handed quarks (and squarks) respectively, the superscript $c$
denotes charge conjugation, $i$, $j$, and $k$ are generation indices and
$\alpha$, $\beta$, and $\gamma$ are SU(3)$_c$ triplet indices.  It follows from
the antisymmetry of $\epsilon^{\alpha\beta\gamma}$ that $\lambda''_{ijk}$ is
antisymmetric in the last two indices, thus there are 9 couplings for three
generations.  The superpotential can be written as
\begin{equation}
W_{UDD} = 2 \epsilon^{\alpha\beta\gamma} \lambda''_{ijk} U^c_{i\alpha}
D^c_{j\beta} D^c_{k\gamma},
\end{equation}
with $k>j$.  The Lagrangian that follows from the above superpotential contains
\begin{eqnarray}
&&{\cal L}_{UDD} \ni - 2\epsilon^{\alpha\beta\gamma} \lambda''_{ijk}
\Big[ \tilde{u}_{Ri\alpha} \overline{\eta}^D_{j\beta}
\overline{\eta}^D_{k\gamma} \nonumber \\
&&\left.+ \tilde{d}_{Rj\beta} \overline{\eta}^U_{i\alpha}
\overline{\eta}^D_{k\gamma} + \tilde{d}_{Rk\gamma} \overline{\eta}^U_{i\alpha}
\overline{\eta}^D_{j\beta} \right]+ {\rm h.c.} ,
\end{eqnarray}
where $\overline{\eta}^D_{i\alpha}$ is the right-handed component of the
down-quark with generation index $i$ and color index $\alpha$, and
$\tilde{d}_{Ri\alpha}$ is the squark associated to it.  To pass to
four-component notation we use the relation $\overline{\eta}_1
\overline{\eta}_2 = \overline{\psi^c_1} P_R \psi_2$, with $P_R = \frac{1}{2}
(1+\gamma_5)$. We obtain
\begin{eqnarray}
&&{\cal L}_{UDD} \ni - 2\epsilon^{\alpha\beta\gamma} \lambda''_{ijk}
\left[ \tilde{u}_{Ri\alpha} \overline{d^c_{j\beta}} P_R d_{k\gamma}\right.
\nonumber \\
&&\left.+ \tilde{d}_{Rj\beta} \overline{u^c_{i\alpha}} P_R
d_{k\gamma} + \tilde{d}_{Rk\gamma} \overline{u^c_{i\alpha}} P_R
d_{j\beta} \right] + {\rm h.c.}
\end{eqnarray}

\begin{figure}[b]
\parbox[b]{1.5in}{FIG. 1.
  Typical diagram for neutralino decay into standard model fermions via
  sfermion exchange.  The R-parity violating vertex is circled.}
\hspace{0.1in}\epsfig{width=1.5in,file=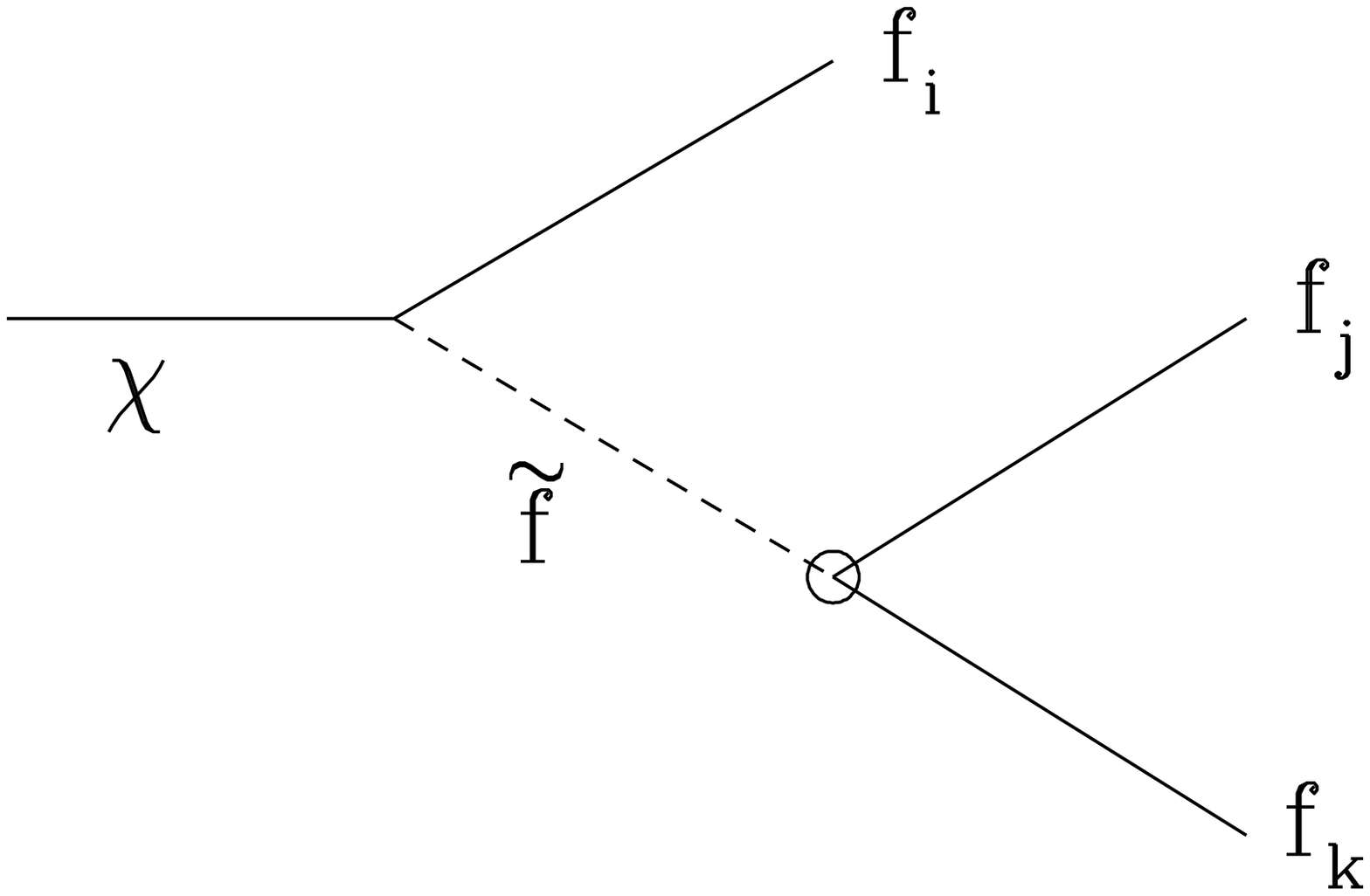}
\end{figure}

The sfermion mass eigenstates $\tilde{f}_\kappa$, with $\kappa=1,\dots,6$, are
related to left- and right-handed sfermions through the $6\times3$ mixing
matrices $\Gamma^f_{R\kappa i}$ and $\Gamma^f_{L\kappa i}$, where $i=1,2,3$ is
a generation index.  Without mixing, $\kappa=i$ is a left-handed sfermion with
generation $i$ and $\kappa=i+3$ is a right handed sfermion with generation $i$.
In general, the sfermion mixings are given by
\begin{equation}
\tilde{f}_\kappa = \Gamma^f_{R\kappa i} \tilde{f}_{Ri} +
\Gamma^f_{L\kappa i} \tilde{f}_{Li}, \label{sfermion1}
\end{equation}
with inverse
\begin{equation}
\tilde{f}_{Ri} = \Gamma^{f*}_{R\kappa i} \tilde{f}_{\kappa}, \qquad
\tilde{f}_{Li} = \Gamma^{f*}_{L\kappa i} \tilde{f}_{\kappa}. \label{sfermion2}
\end{equation}
Note that there is only a left-handed sneutrino, so there is only a left handed
mixing matrix and $\kappa=1,2,3$.  In the simple case of no flavor mixing but
with left-right mixing, the only non-zero elements are
\begin{eqnarray}
\Gamma^f_{Lii} &= \Gamma^f_{R(i+3)i} &= \cos\theta_f, \nonumber \\
\Gamma^f_{Rii} &= - \Gamma^f_{L(i+3)i} &= \sin\theta_f,
\end{eqnarray}
where $\theta_f$ is the left-right mixing angle for sfermion $\tilde{f}$.

Inserting the mixing matrices into the previous Lagrangian, we have
\begin{eqnarray}
&&{\cal L}_{UDD} \ni - 2\epsilon^{\alpha\beta\gamma} \lambda''_{ijk}
\left[ \Gamma^{u*}_{R\kappa i} \tilde{u}_{\kappa\alpha}
\overline{d^c_{j\beta}} P_R d_{k\gamma}\right. \nonumber \\
&&\left.+ \Gamma^{d*}_{R\kappa j}
\tilde{d}_{\kappa\beta} \overline{u^c_{i\alpha}} P_R d_{k\gamma} +
\Gamma^{d*}_{R\kappa k} \tilde{d}_{\kappa\gamma} \overline{u^c_{i\alpha}}
P_R d_{j\beta} \right]+ {\rm h.c.}
\end{eqnarray}
The R-parity violating vertices are read directly from this Lagrangian.

We also need the neutralino-sfermion-fermion vertex. It comes from the
interaction term
\begin{equation}
{\cal L}_{\chi f\tilde{f}} \ni \overline{\chi} (g^L_{\chi f_i\kappa} P_L
+ g^R_{\chi f_i\kappa} P_R ) f_{i\alpha} \tilde{f}^{*}_{\kappa\alpha}+
{\rm h.c.},
\label{chi-sfermion}
\end{equation}
where $i$ is a fermion generation index, $\kappa$ specifies the sfermion mass
eigenstate, and $\alpha$ is a color index. The couplings are
\begin{equation}
\label{eq:coupl1}
g^A_{\chi f_i\kappa} = \Gamma^f_{R\kappa i} g^{RA}_{\chi f_i} + 
                \Gamma^f_{L\kappa i} g^{LA}_{\chi f_i}, \\
\end{equation}
where $A$ can be $L$ or $R$ and
\begin{eqnarray}
g^{LL}_{\chi f_i} &=& -\sqrt{2} \left[ \left(q_f-T_3\right) g'
N_{\chi 1} + T_3 g N_{\chi 2} \right], \\
g^{LR}_{\chi u_i} &=& g^{RL}_{\chi u_i} = -\frac{ g m_{u_i} N_{\chi 4} } 
     {\sqrt{2} m_W \sin\beta}, \\
g^{LR}_{\chi d_i} &=& g^{RL}_{\chi d_i} = -\frac{ g m_{d_i} N_{\chi 3} } 
     {\sqrt{2} m_W \cos\beta}, \\
g^{RR}_{\chi f_i} &=& +\sqrt{2} q_f g' N_{\chi 1}.
\label{eq:coupl2}
\end{eqnarray}
Here $T_3$ is the third component of the weak isospin, $N_{ij}$ is the
$4\times4$ neutralino mixing matrix in the convention in which all neutralino
masses are positive, $u$ is an up-type quark or neutrino $(T_3=+1/2)$, and $d$
is a down-type quark or charged lepton $(T_3=-1/2)$.  The charges are $q_u =
2/3$, $q_d=-1/3$, $q_\nu=0$, and $q_e=-1$.  Notice that the only surviving
neutrino couplings are $g^{LL}_{\chi\nu_i}$.

We can now obtain the full expression for the decay amplitude.  In order to get
correct signs for the interference terms we use Wick's theorem.  The effective
operator for neutralino decay is
\begin{eqnarray}
{\cal T}(\chi \to \ubar_i \dbar_j \dbar_k) =&& \\
 -2\epsilon^{\alpha\beta\gamma} \lambda''_{ijk} \Big[&&
\overline{\chi} (G^{RL}_{u_i}P_L+G^{RR}_{u_i}P_R) u_{i\alpha}
\overline{d^c_{j\beta}} P_R d_{k\gamma} + \nonumber \\ &&
\overline{\chi} (G^{RL}_{d_j}P_L+G^{RR}_{d_j}P_R) d_{j\beta}
\overline{u^c_{i\alpha}} P_R d_{k\gamma} + \nonumber \\ &&
\overline{\chi} (G^{RL}_{d_k}P_L+G^{RR}_{d_k}P_R) d_{k\gamma}
\overline{u^c_{i\alpha}} P_R d_{j\beta} \Big], \nonumber
\end{eqnarray}
where
\begin{equation}
G^{AB}_{f_i} = \sum_{\kappa} \Gamma^{f*}_{A\kappa i}
\Delta^*_{\tilde{f}_\kappa} g^{B}_{\chi f_i\kappa},
\end{equation}
$A$ and $B$ are either $R$ or $L$ and
$\Delta_{\tilde{f}_\kappa}=(p^2-m_{\tilde{f}_\kappa}^2+
im_{\tilde{f}_\kappa}\Gamma_{\tilde{f}_\kappa})^{-1}$ is the sfermion
propagator.  In the case of no sfermion mixing, $\theta_f=0$, we find
\begin{equation}
G^{AB}_{f_i} = \Delta^*_{\tilde{f}_{Ai}} g^{AB}_{\chi f_i}.
\end{equation}
In all cases $G^*$ corresponds to a fermion and $G$ corresponds to an
antifermion in the decay amplitudes.  The rates for the charge conjugated
decays are found by complex conjugating $G$ and $G^*$ everywhere in our
expressions for the spin summed squared amplitudes and thus are identical.

Suppressing color wavefunctions, we obtain the following amplitude for the
decay $\chi \to \ubar_i\dbar_j\dbar_k$:
\begin{eqnarray}
{\cal M}(\chi \to \ubar_i\dbar_j&&\dbar_k) =
\langle \ubar_i\dbar_j\dbar_k | {\cal T} | \chi
\rangle \\
= - 2\lambda''_{ijk} \Big[&&
\overline{v}_\chi (G^{RL}_{u_i}P_L+G^{RR}_{u_i}P_R) v_{u_i}
\overline{u}_{d_j} P_R v_{d_k} - \nonumber \\ &&
\overline{v}_\chi (G^{RL}_{d_j}P_L+G^{RR}_{d_j}P_R) v_{d_j}
\overline{u}_{u_i} P_R v_{d_k} + \nonumber \\ &&
\overline{v}_\chi (G^{RL}_{d_k}P_L+G^{RR}_{d_k}P_R) v_{d_k}
\overline{u}_{u_i} P_R v_{d_j} \Big], \nonumber
\end{eqnarray} 
where $u_f$ and $v_f$ are the usual particle and antiparticle Dirac spinors
associated with quark $f$.  Notice the minus sign in the second term in
brackets, which comes from Dirac statistics. It is comforting that the same
relative sign is obtained from the familiar Feynman rule of the sign of the
fermion permutations: the first term has the fermions in the order
$(\chi,u_i,d_j,d_k)$, the second in the order $(\chi,d_j,u_i,d_k)$, and the
third in the order $(\chi,d_k,u_i,d_j)$. The second and the third are odd and
even permutations of the first. This confirms the relative minus sign for the
second term.

Squaring the amplitude, averaging over the two spin states of the neutralino,
and summing over the final antiquark spin states, we obtain
\begin{eqnarray}
\label{eq:Muds}
&&{\sum_{\rm spins}}' \Big| {\cal M}(\chi \to
\ubar_i\dbar_j\dbar_k) \Big|^2 =
8 c_f |\lambda^{\prime\prime}_{ijk}|^2 \times \Re\Big\{ \\
&&d_j \cdot d_k\Big[(|G^{RL}_{u}|^2+|G^{RR}_{u}|^2) \chi \cdot u
+ 2 G^{RL}_{u} G^{RR*}_{u} m_\chi m_{u}\Big] \nonumber \\
&&+ u \cdot d_k\left[(|G^{RL}_{d_j}|^2+|G^{RR}_{d_j}|^2)\chi \cdot d_j 
+ 2 G^{RL}_{d_j} G^{RR*}_{d_j} m_\chi m_{d_j}\right] \nonumber \\
&&+ u \cdot d_j\Big[(|G^{RL}_{d_k}|^2+|G^{RR}_{d_k}|^2) \chi \cdot d_k
+ 2 G^{RL}_{d_k} G^{RR*}_{d_k} m_\chi m_{d_k}\Big] \nonumber \\
&&- G^{RR}_{u} G^{RR*}_{d_j} g(u,d_k,d_j,\chi)
- G^{RL}_{u} G^{RR*}_{d_j}(d_k \cdot d_j)m_\chi m_{u} \nonumber \\
&&- G^{RR}_{u} G^{RL*}_{d_j}(d_k \cdot u)m_\chi m_{d_j}
- G^{RL}_{u} G^{RL*}_{d_j}(\chi \cdot d_k) m_{u} m_{d_j}
\nonumber \\
&&- G^{RR}_{u} G^{RR*}_{d_k} g(u,d_j,d_k,\chi) 
- G^{RL}_{u} G^{RR*}_{d_k}(d_k \cdot d_j)m_\chi m_{u} \nonumber \\
&&- G^{RR}_{u} G^{RL*}_{d_k}(u \cdot d_j) m_\chi m_{d_k}
- G^{RL}_{u} G^{RL*}_{d_k}(\chi\cdot d_j) m_{u} m_{d_k}
\nonumber \\
&&- G^{RR}_{d_j} G^{RR*}_{d_k} g(d_j,u,d_k,\chi) 
- G^{RL}_{d_j} G^{RR*}_{d_k}(d_k \cdot u)m_\chi m_{d_j}\nonumber \\
&&- G^{RR}_{d_j} G^{RL*}_{d_k}(d_j \cdot u)m_\chi m_{d_k}
- G^{RL}_{d_j} G^{RL*}_{d_k}(\chi \cdot u)m_{d_j}m_{d_k} \Big\},
\nonumber
\end{eqnarray}
with the color factor $c_f=6$, and $g(a,b,c,d) = (a \cdot b) (c \cdot d) - (a
\cdot c) (b \cdot d) + (a \cdot d) (b \cdot c)$. Particle four-momenta have
been denoted by the particle letter and unambiguous indices have been
suppressed.

The calculation of the $LQD^c$ decays is very similar.  The superpotential is
\begin{equation} 
  W_{LQD} = \epsilon^{\sigma\rho}\left[\lambda'_{ijk}
  L_{i\sigma} Q_{j\rho\alpha} D^c_{k\alpha}\right],
\end{equation}
where $\sigma$ and $\rho$ are SU(2)$_L$ indices and $\alpha$ is an SU(3)$_c$
index.  Suppressing color wavefunctions, we obtain the Lagrangian
\begin{eqnarray}
&& {\cal L}_{LQD} \ni \lambda'_{ijk} \left[ \tilde{e}_{Li}
\overline{d_{k}} P_L u_{j} + \tilde{u}_{Lj} \overline{d_{k}} P_L e_i +
\tilde{d}_{Rk}^* \overline{e^c_i} P_L u_{j} \right. \nonumber \\ &&
\left. - \tilde{\nu}_{Li} \overline{d_{k}} P_L d_{j} -
\tilde{d}_{Lj} \overline{d_{k}} P_L \nu_i -
\tilde{d}_{Rk}^* \overline{\nu^c_i} P_L d_{j} \right] + {\rm h.c.}
\end{eqnarray}
We have used the identities $\xi_1 \eta_2 = \overline{\psi_2} P_L \psi_1$ and
$\xi_1 \xi_2 = \overline{\psi^c_1} P_L \psi_2$ to pass to four-component
notation.  Here there are 27 couplings for three generations as the coupling
matrix is unconstrained by symmetry arguments.

After introducing sfermion mass eigenstates through Eq.~(\ref{sfermion2}),
multiplying by Eq.~(\ref{chi-sfermion}), and using Wick's theorem, we obtain
the following amplitudes for the decays $\chi \to e^+_i\ubar_jd_k$ and
$\chi\to\nubar_i\dbar_jd_k$:
\begin{eqnarray}
{\cal M}(\chi \to& e^+_i\ubar_jd_k) &= \\
&- \lambda'_{ijk} \Big[
&\overline{v}_\chi (G^{LL}_{e_i}P_L+G^{LR}_{e_i}P_R) v_{e_i}
\overline{u}_{d_k} P_L v_{u_j} - \nonumber \\ &&
\overline{v}_\chi (G^{LL}_{u_j}P_L+G^{LR}_{u_j}P_R) v_{u_j}
\overline{u}_{d_k} P_L v_{e_i} + \nonumber \\ &&
\overline{u}_{d_k} (G^{RR*}_{d_k}P_L+G^{RL*}_{d_k}P_R) u_\chi
\overline{u}_{e_i} P_L v_{u_j} \Big], \nonumber
\end{eqnarray} 
\begin{eqnarray}
{\cal M}(\chi \to& \nubar_i\dbar_jd_k) &= \\
&\lambda'_{ijk} \Big[
&\overline{v}_\chi G^{LL}_{\nu_i}P_L v_{\nu_i}
\overline{u}_d P_L v_\dbar - \nonumber \\ &&
\overline{v}_\chi (G^{LL}_{d_j}P_L+G^{LR}_{d_j}P_R) v_{d_j}
\overline{u}_{d_k} P_L v_{\nu_i} + \nonumber \\ &&
\overline{u}_{d_k} (G^{RR*}_{d_k}P_L+G^{RL*}_{d_k}P_R) u_\chi
\overline{u}_{\nu_i} P_L v_{d_j} \Big]. \nonumber
\end{eqnarray}
The matrix elements squared then follow as
\begin{eqnarray}
\label{eq:Mude}
&&{\sum_{\rm spins}}' \Big| {\cal M}(\chi \to
e^+_i\ubar_jd_k) \Big|^2 =2 c_f |\lambda^\prime_{ijk}|^2
\times\Re\Big\{ \\
&&e \cdot d\left[ (|G^{LL}_u|^2+|G^{LR}_u|^2) \chi \cdot u
+ 2 G^{LL}_u G^{LR*}_u m_\chi m_u\right] \nonumber \\
&&+ u \cdot e \left[(|G^{RL*}_d|^2+|G^{RR*}_d|^2) \chi \cdot d 
+ 2 G^{RL*}_d G^{RR}_d m_\chi m_d \right] \nonumber \\
&&+ u \cdot d \left[ (|G^{LL}_e|^2+|G^{LR}_e|^2) \chi \cdot e
+ 2 G^{LL}_e G^{LR*}_e m_\chi m_e\right] \nonumber \\
&&- G^{LL}_u G^{RR}_d g(u,e,d,\chi) 
- G^{LR}_u G^{RR}_d (e \cdot d) m_\chi m_u \nonumber \\
&&- G^{LL}_u G^{RL}_d (e \cdot u) m_\chi m_d
- G^{LR}_u G^{RL}_d (\chi \cdot e) m_u m_d \nonumber \\
&&- G^{LL}_u G^{LL*}_e g(e,d,u,\chi) 
- G^{LR}_u G^{LL*}_e (e \cdot d) m_\chi m_u \nonumber \\
&&- G^{LL}_u G^{LR*}_e (u \cdot d) m_\chi m_e
- G^{LR}_u G^{LR*}_e (\chi \cdot d) m_u m_e \nonumber \\
&&- G^{RR*}_d G^{LL*}_e g(e,u,d,\chi) 
- G^{RL*}_d G^{LL*}_e (e \cdot u) m_\chi m_d \nonumber \\
&& - G^{RR*}_d G^{LR*}_e (d \cdot u) m_\chi m_e
- G^{RL*}_d G^{LR*}_e (\chi \cdot u) m_d m_e \Big\}, \nonumber 
\end{eqnarray}
\begin{eqnarray}
\label{eq:Mddnu}
&&{\sum_{\rm spins}}' \Big| {\cal M}(\chi \to
\nubar_i\dbar_jd_k) \Big|^2 = 2 c_f |\lambda^\prime_{ijk}|^2 
\times \Re\Big\{\\
&&\nu \cdot d \left[(|G^{LL}_{d_j}|^2+|G^{LR}_{d_j}|^2)\chi \cdot \dbar
+ 2 G^{LL}_{d_j} G^{LR*}_{d_j} m_\chi m_{d_j}\right]\nonumber \\
&&+ \dbar \cdot \nu \left[(|G^{RL*}_{d_k}|^2+|G^{RR*}_{d_k}|^2) 
\chi \cdot d + 2 G^{RL*}_{d_k} G^{RR}_{d_k} m_\chi m_{d_k} \right]
\nonumber \\
&&+ \dbar \cdot d |G^{LL}_\nu|^2 \chi \cdot \nu \nonumber \\
&&- G^{LL}_{d_j} G^{RR}_{d_k} g(\dbar,\nu,d,\chi)
- G^{LR}_{d_j} G^{RR}_{d_k} (\nu \cdot d) m_\chi m_{d_j} \nonumber \\
&&- G^{LL}_{d_j} G^{RL}_{d_k} (\nu \cdot \dbar) m_\chi m_{d_k} 
- G^{LR}_{d_j} G^{RL}_{d_k} (\chi \cdot \nu) m_{d_j} m_{d_k} \nonumber \\
&&- G^{LL}_{d_j} G^{LL*}_\nu g(\nu,d,\dbar,\chi)
- G^{LR}_{d_j} G^{LL*}_\nu (\nu \cdot d) m_\chi m_{d_j} \nonumber \\
&&- G^{RR*}_{d_k} G^{LL*}_\nu g(\nu,\dbar,d,\chi)
- G^{RL*}_{d_k} G^{LL*}_\nu (\nu \cdot \dbar) m_\chi m_{d_k} \Big\}, \nonumber
\end{eqnarray}
where now the color factor is given by $c_f=3$ in both.  Note that in
Eqs.~(\ref{eq:Mude}) and~(\ref{eq:Mddnu}) for the $LQD^c$ processes the factor
in front is 2, not 8 as in Eq.~(\ref{eq:Muds}) for the $U^cD^cD^c$ processes.
This is because the $U^cD^cD^c$ process is identical for $\lambda''_{ijk}$ and
$\lambda''_{ikj}$, whereas each element of the $LQD^c$ matrix $\lambda'_{ijk}$
gives a unique channel.

The calculation of the $LLE^c$ decay is very similar to the $LQD^c$ case.  From
the superpotential
\begin{equation}
W_{LLE} = \epsilon^{\sigma\rho}\left[\lambda_{ijk} L_{i\sigma} L_{j\rho}
E^c_k\right],
\end{equation} 
we obtain the Lagrangian
\begin{eqnarray}
&& {\cal L}_{LLE} \ni 2\lambda_{ijk} \left[ \tilde{e}_{Li} \overline{e_k} P_L
\nu_j + \tilde{\nu}_{Lj} \overline{e_k} P_L e_i + \tilde{e}_{Rk}^*
\overline{e^c_i} P_L \nu_j \right. \nonumber \\ &&
\left. -\tilde{\nu}_{Li} \overline{e_k} P_L e_j -
\tilde{e}_{Lj} \overline{e_k} P_L \nu_i - \tilde{e}_{Rk}^* \overline{e^c_j}
P_L \nu_i \right] + {\rm h.c.},
\end{eqnarray}
where $j>i$ by the same antisymmetry argument as in the $U^cD^cD^c$ case.
Again there are 9 couplings for three generations.

Passing to sfermion mass eigenstates and using Wick's theorem, we obtain the
following amplitude for the decay $\chi \to e^+_i\nubar_j e^-_k$:
\begin{eqnarray}
{\cal M}(\chi \to& e^+_i\nubar_j e^-_k) &= \\
&- 2\lambda_{ijk} \Big[
&\overline{v}_\chi (G^{LL}_{e_i}P_L+G^{LR}_{e_i}P_R) v_{e_i}
\overline{u}_{e_k} P_L v_{\nu_j} - \nonumber \\ &&
\overline{v}_\chi G^{LL}_{\nu_j}P_L v_{\nu_j}
\overline{u}_{e_k} P_L v_{e_i} + \nonumber \\ &&
\overline{u}_{e_k} (G^{RR*}_{e_k}P_L+G^{RL*}_{e_k}P_R) u_\chi
\overline{u}_{e_i} P_L v_{\nu_j} \Big]. \nonumber
\end{eqnarray}
The matrix element squared is
\begin{eqnarray}
\label{eq:Mlle}
&&{\sum_{\rm spins}}' \Big| {\cal M}(\chi \to
e^+_i\nubar_j e^-_k) \Big|^2=8 |\lambda_{ijk}|^2
\times \Re \Big\{ \\
&&e \cdot \nu\left[ (|G^{LL}_{e_i}|^2+|G^{LR}_{e_i}|^2) \chi \cdot \ebar
+ 2 G^{LL}_{e_i} G^{LR*}_{e_i} m_\chi m_{e_i}\right] \nonumber \\
&&+ e \cdot \ebar |G^{LL}_\nu|^2\chi \cdot \nu \nonumber \\
&&+ \ebar \cdot \nu\left[ (|G^{RL*}_{e_k}|^2+|G^{RR*}_{e_k}|^2)
\chi \cdot e + 2 G^{RL*}_{e_k} G^{RR}_{e_k} m_\chi m_{e_k}\right]
\nonumber \\
&&- G^{LL}_{e_i} G^{LL*}_\nu g(\ebar,e,\nu,\chi) 
- G^{LR}_{e_i} G^{LL*}_\nu (e \cdot \nu) m_\chi m_{e_i} \nonumber \\
&&- G^{LL}_{e_i} G^{RR}_{e_k} g(\ebar,\nu,e,\chi)
- G^{LR}_{e_i} G^{RL}_{e_k} (\chi \cdot \nu) m_{e_k} m_{e_i} \nonumber \\
&&- G^{LR}_{e_i} G^{RR}_{e_k} (e \cdot \nu) m_\chi m_{e_i}
- G^{LL}_{e_i} G^{RL}_{e_k} (\ebar \cdot \nu ) m_\chi m_{e_k}  \nonumber \\
&&- G^{LL}_\nu G^{RR}_{e_k} g(\nu,\ebar,e,\chi)
- G^{LL}_\nu G^{RL}_{e_k} (\nu \cdot \ebar) m_\chi m_{e_k}\Big\}. \nonumber
\end{eqnarray}
Again, there is a factor of 8 instead of 2 due to the fact that $\lambda_{ijk}$
and $\lambda_{jik}$ allow the same process.  Note that if $i>j$, the amplitude
gains an overall minus sign, but the squared amplitude is identical.

In the limit of no sfermion mixing, we can compare our results with those of
Butterworth, Dreiner, and Morawitz \cite{dreiner}.  We differ in several
respects. In all three types of decays we keep complex conjugations in the
couplings, which are crucial when the neutralino mass eigenvalue is negative,
and we do not have the global phase space factor of $2(1-m^2_1/E^2_1)^{-1/2}$
which appears in their rates.  Our $LLE^c$ and $U^cD^cD^c$ decay rates are a
factor of four larger than theirs, due to the fact that each channel is
duplicated in the coupling matrix which doubles the decay amplitude.  Finally,
we find that in the $U^cD^cD^c$ decay their couplings $a$ and $b$ are
exchanged, in other words gauginos and higgsinos are exchanged, and that in
some instances, particles and antiparticles have been confused (because the
$a$'s are the same for particles and antiparticles but the $b$'s differ).

Together with H. Dreiner \cite{private}, we have agreed that the formulas in
this paper are the correct ones.  Erratum to \cite{dreiner} will be published
elsewhere \cite{private}.

In conclusion, we have computed the neutralino decay rate in R-parity violating
extensions to the minimal supersymmetric standard model.  For the first time we
include complete mixing among sfermions.  Our results supersede previously
published calculations \cite{dreiner}.

We thank H. Dreiner for taking the time to compare with our results and for
comments on the manuscript.  This research was supported by grants from NASA
and DOE.

\setlength{\parindent}{0cm}

\setlength{\baselineskip}{.5\baselineskip}

\end{document}